\documentclass{article}
\usepackage{latexsym,amsmath,amssymb}

\renewcommand{\l}{\langle}
\renewcommand{\r}{\rangle}
\renewcommand{\phi}{\varphi}

\def\mv#1{\xrightarrow{#1}}

\newcommand{\cfg}{C \! f \! g}

\newcounter{tbsnr}
\newenvironment{tbs}
{\addtocounter{tbsnr}{1}\par\bigskip \noindent\fbox{\thetbsnr}
\hspace*{\fill}\begin{minipage}{10cm}\tt}
{\end{minipage}\hspace*{\fill}\bigskip}

\newenvironment{proof}{{\em Proof.}}{\hfill $\Box$ \\}

\newtheorem{theorem}{Theorem}
\newtheorem{lemma}[theorem]{Lemma}

\begin{document}

\title{Programming and Verifying \\  Subgame-Perfect Mechanisms}

\author{Marc Pauly \vspace{0.3 cm} \\
\small CNRS, IRIT \\
\small \texttt{pauly@irit.fr}}

\date{\today}

\maketitle

\begin{abstract}
An extension of the WHILE-language is developed for programming
game-theoretic mechanisms involving multiple agents. Examples of
such mechanisms include auctions, voting procedures, and
negotiation protocols. A structured operational semantics is
provided in terms of extensive games of almost perfect
information. Hoare-style partial correctness assertions are
proposed to reason about the correctness of these mechanisms,
where correctness is interpreted as the existence of a subgame-perfect 
equilibrium. Using an extensional approach to pre- and
postconditions, we show that an extension of Hoare's original
calculus is sound and complete for reasoning about subgame-perfect
equilibria in game-theoretic mechanisms.  We use the calculus to
verify some simple mechanisms like the Dutch auction.
\end{abstract}

%
%
\section{Introduction}
%
%

In recent years, games have become more prominent in different
areas of computer science research. The reason for this seems to
be the realisation that games form a natural generalisation of
programs. This insight can be realised on a number of different
levels (we shall only mention two): On a foundational level, games
have been used to provide an alternative model of computation, the
alternating Turing machine \cite{alternating-TM}. At a more
abstract level, program logics like propositional dynamic logic
have been extended to games \cite{Parikh}.

From a game-theoretic perspective, much of this work is extremely
narrow, since it mainly focuses on determined 2-player win/loss
games of perfect information. On the other hand, game theory has
developed a wealth of techniques to study more complicated
situations where agents interact, involving more than two players,
imperfect information, and preferences over outcomes which cannot
be captured by simply distinguishing between winning and losing.
Still, it has been suggested \cite{Parikh:SoSo} that combining
research in game theory and computer science, we may be able to
obtain a better understanding of {\em social software}, i.e., of
the formal properties of the social processes we are involved in.
The present paper tries to contribute to this aim.

More concretely, we attempt to generalise techniques from formal
program verification to games or game-theoretic mechanisms such as
auctions, voting procedures, etc. From a logical perspective, two
approaches suggest themselves. On the one hand, one might extend
model checking approaches \cite{Clarke}, where one uses, for
instance, temporal logic to specify properties of a system
(program/game) and proceeds to verify these properties using model
checking. This approach has been generalised to reason about
coalitional power in games \cite{coal1}. On the other hand, one
can try to extend approaches based on theorem proving using a
formal calculus in which one can derive certain properties of a
system. This approach will be taken here.

The axiomatic or compositional approach to program verification
was introduced by Hoare \cite{Hoare-logic} and Dijkstra \cite{D},
and provided the foundation stone for formal program verification
\cite{Nielson,Apt-Olderog,Francez}. In Hoare's calculus,
correctness assertions of the form $\{P\} \pi \{Q\}$ are used to
express that program $\pi$, when executed in a state satisfying
$P$, will terminate in a state satisfying $Q$ (provided it does
terminate). In generalising the
program verification approach to games, this paper makes two
contributions: First, it defines a programming language which is a
simple extension of the WHILE-language sufficient to program
game-theoretic mechanisms. The syntax of this language is defined
in section \ref{syntax}, and section \ref{semantics} provides a
structured operational semantics in terms of extensive games of
almost perfect information. Second, we are going to extend Hoare's
calculus to reason about the correctness of these mechanisms,
where correctness is interpreted as the existence of a subgame-perfect 
equilibrium with a certain payoff. In section
\ref{mech-des}, we define our new notion of correctness by
providing a game-theoretic interpretation of $\{P\}\pi\{Q\}$, also
linking it to the game-theoretic notion of implementation and
mechanism design.  Section \ref{calculus} presents an extensional
calculus for reasoning about mechanism correctness, and provides
proofs of soundness and completeness.  Finally, section \ref{apps}
illustrates the calculus in the verification of a few simple mechanisms.

%
%
\section{Syntax of MPL}
\label{syntax}
%
%

Our mechanism programming language (MPL) is a simple extension of
standard imperative programming languages; more concretely,
our point of departure is the well-known WHILE-language (see e.g.
\cite{Nielson}).
We assume throughout that we are given a nonempty set of agents or players
$Ags$,
a set of mechanism variables $MV$, a set of function symbols $Funs$ and a set
of relation symbols $Rels$. Using these, we inductively define
terms $t$, boolean expressions $B$ and
mechanisms (or game forms) $\gamma$ as follows:

\[ \begin{array}{|rl|}
\hline
t := & x \; | \; f^k(t_1, \ldots, t_k) \\[1ex]

B := & true \; | \; R^k(t_1, \ldots, t_k) \; | \;
 \neg B \; | \; B_{1} \wedge B_{2} \\[1ex]

\gamma := & x:=t \; | \; \gamma_{1} ; \gamma_{2} \; | \;
\mathtt{if} \: B \: \mathtt{then} \: \gamma_{1} \: \mathtt{else} \: \gamma_{2}
\; | \; \mathtt{while} \: B \: \mathtt{do} \: \gamma \; | \; \\
& \mathtt{ch}_A(\{x_a | a \in A\}) \\ \hline
\end{array} \]
where $a {\in} Ags$, $f^k {\in} Funs$ is a $k$-ary function symbol
(in case $k=0$ we are dealing
with constants), $R^k {\in} Rels$ a $k$-ary
relation symbol, $x,x_a {\in} MV$, and $A \subseteq Ags$ is finite and
nonempty.

The last construct presents the only addition to the standard
WHILE-language: $\mathtt{ch}_A$ lets agent $a \in A$ choose any value for
the variable $x_a$.  The agents in $A$ are making their choice
simultaneously, so in order to prevent conflicting assignments to
variables, we require all the $x_a$ to be distinct.  One can think of
the $\mathtt{ch}_A$ construct as a strategic game among $n$ agents,
where the strategic choice of an agent is represented by the value of
his/her variable.  While the set of agents may be infinite, we require
each $\mathtt{ch}_A$ construct to involve only finitely many agents. 
In the special case where $|A|=1$, we have a simple nondeterministic
choice.  More concretely, in case agent $1$ can choose between 2
different strategies, executing $\gamma_{1}$ vs.\ executing
$\gamma_{2}$, we can describe this situation as \[
\mathtt{ch}_{\{1\}}(\{x_1\}) ; \mathtt{if} \: x_1=0 \: \mathtt{then}
\: \gamma_{1} \: \mathtt{else} \: \gamma_{2} ,\] where we assume that
the domain of computation is the set of natural numbers, for instance,
and $= \: \in Rels$ and $0 \in Funs$.

MPL is an extremely general programming language for a large
variety of different kinds of mechanisms. In section
\ref{mech-des} we shall use it for defining mechanisms for
different kinds of auctions. Voting procedures are further
examples of mechanisms which can be programmed using MPL. As an
example, the well-known Borda-count procedure (see, e.g.,
\cite{Brams:voting}) can be programmed as follows:
\begin{tabbing}
$\mathtt{ch}_{Ags}(\{x_1,x_2, \ldots, x_N\})$; \\
$i:=1;$ \\
$\mathtt{while} \: i \leq K \: \mathtt{do} \: c_i:=0; i:=i+1$; \\
$a:=1;$ \\
$\mathtt{while}$ \= $\: a \leq N \: \mathtt{do}$ \\
    \> $i:=1;$ \\
    \> $\mathtt{while}$ \= $\: i \leq K \: \mathtt{do}$ \\
    \> \> $c_i:= c_i + x_a[i];$ \\
    \> \> $i:=i+1;$ \\
    \> $a:=a+1$
\end{tabbing}
In this example, we assume that $Ags=\{1, 2, \ldots, N\}$, and
that the agents have to choose among $K$ candidates. First, each
agent $a$ can cast a ballot of the form $x_a = (p_1, p_2, \ldots,
p_K)$, where $p_i$ is the number of points the agent gives to
candidate $i$. Ballots have to be rankings of candidates, i.e.,
the most preferred candidate must obtain $K$ points, the next
preferred candidate $K-1$ points, etc., so that the least-preferred 
candidate obtains 1 point. Hence, we assume implicitly
that the domain of computation contains these possible ballots,
and that the initial choice assigns a ballot to each $x_a$. (Note
that since the domain of computation will also contain the natural
numbers, we need to make sure that each $x_a$ is assigned to an
element of the appropriate ballot type, but we shall ignore this
problem in order to keep the algorithm simple.) Once the ballots
are cast, $a$ is initialised to the first agent, and $i$ to the
first candidate. The variable $c_i$ counts the number of points
accumulated by candidate $i$, and is initialised to $0$. The main
part of the algorithm then simply sums up the points for each
candidate, where $x_a[i]$ refers to $p_i$, in case $x_a = (p_1,
p_2, \ldots, p_K)$. The winner of the vote will be the candidate
accumulating the most points.

A further example of a well-known mechanism which can be
programmed in MPL is a version of Rubinstein's negotiation
protocol of alternating offers (see \cite{Osborne,Kraus}).
\begin{tabbing}
$agree:=false;$ \\
$optout:=false;$ \\
$i:=1$; \\
$\mathtt{while}$ \= $\: \neg optout \wedge \neg agree \: \mathtt{do}$ \\
\> $\mathtt{if} \: i=1 \: \mathtt{then} \:
\mathtt{ch}_{\{1\}}(\{x\}) \mathtt{else} \:
\mathtt{ch}_{\{2\}}(\{x\});$ \\
\> $\mathtt{if} \: i=1 \: \mathtt{then} \:
\mathtt{ch}_{\{2\}}(\{y\}) \mathtt{else} \:
\mathtt{ch}_{\{1\}}(\{y\});$ \\
\> $\mathtt{if} \: y=0 \:$ \= $\mathtt{then} \: agree:= true$ \\
\>\> $\mathtt{else} \: \mathtt{if} \: y=1 \: \mathtt{then} \:
optout:= true \: \mathtt{else} \: i:=3-i$
\end{tabbing}
For simplicity, we have assumed that there are only two agents who
try to reach an agreement over, e.g., the price of a car which
agent 1 wants to sell to agent 2, and so we can assume the domain
of computation to be simply the natural numbers. The negotiation
procedure can end in an agreement concerning the price, one of the
agents can opt out of the negotiation (in which case some
predetermined event will occur), or the negotiation can go on
forever. The protocol starts by agent 1 making a price offer $x$.
Agent 2 responds by choosing $y$, where we interpret $y=0$ as
signalling agreement to the price offered, $y=1$ as a decision to
opt out of the negotiation, and any other value for $y$ as
signalling the desire to make a counteroffer, upon which we get
another iteration of the loop with the roles reversed.

The above negotiation protocol is very general, and numerous
instances of it have been analysed game-theoretically
\cite{Kraus}. We shall not go into this or the voting mechanism in
more detail, since our main aim at this point is only to suggest the 
generality of the mechanism programming language defined.  Section
\ref{apps} will provide a more detailed and more formal treatment of
examples such as the ones given here. In the following section, we shall
provide a formal semantics for this language in terms of games. 
Furthermore, we will subsequently provide a calculus for reasoning
about the existence of game-theoretic equilibria in these mechanisms,
and about the payoffs the agents obtain in equilibrium.

Note that MPL only allows one to construct mechanisms with
almost-perfect information, i.e., agents are perfectly informed
about all the choices made except possibly for simultaneous moves.
Different subclasses of MPL-mechanisms correspond to various
natural assumptions regarding the power of the mechanism designer
and the agents in general.  The class MPL(PRG) of {\em programs}
is the class of MPL-mechanisms which do not contain any
$\mathtt{ch}_A$ construct. Without this construct, MPL is simply
the WHILE-language. The class MPL(PI) of {\em perfect-information
mechanisms} will restrict the use of $\mathtt{ch}_A$ to cases
where $|A|=1$, i.e., where all choices involve only a single
agent. Perfect-information mechanisms allow different agents to
make choices at different times, but all choices are public, there are 
no simultaneous moves.

%
%
\section{Structured Operational Semantics via Games}
\label{semantics}
%
%

The most detailed semantics we can provide for MPL expressions is a
structured operational semantics which specifies the configurations a
mechanism can be in and the possible transitions between
configurations.  For programs, such a semantics gives rise to an
execution sequence or trace, and in case of nondeterministic programs to
an execution tree.  Since in the case of mechanisms we are dealing with
multiple agents, we arrive at a game tree whose
positions are the possible configurations of the mechanism.

As is standard in first-order logic, we will work with an
interpretation ${\mathcal I}$ which provides us with a domain
$D_{\mathcal I}$ and functions and relations over $D_{\mathcal I}$
as interpretations for the symbols in $Funs$ and $Rels$.
Furthermore, we assume that besides the relations associated to
symbols in $Rels$, our interpretation contains an additional
binary $\geq_a^{\mathcal I}$-relation for every agent $a \in Ags$.
The $\geq_a^{\mathcal I}$ relation will be used to represent agent
$a$'s preference over the elements of the domain. Note that
mechanisms programmed in MPL cannot refer to these preferences,
since $\geq_a \not\in Rels$.

The only requirements on ${\mathcal I}$ are that the preference
relations $\geq_a^{\mathcal I} \subseteq D_{\mathcal I} \times
D_{\mathcal I}$ satisfy the following properties: (1)
$\geq_a^{\mathcal I}$ must be a partial pre-order, i.e., a
reflexive and transitive relation on $D_{\mathcal I}$, and (2)
there is a uniformly worst outcome (which we denote as $-\infty$),
i.e., there is some $d \in D_{\mathcal I}$ such that for all $a
\in Ags$ and $x \in D_{\mathcal I}$ we have $x \geq_a^{\mathcal I} d$. 
Usually, preference relations will be total orders, but our framework
does not require this. The uniformly worst outcome is needed to deal
with some infinite runs resulting from while-loops, it plays no
substantive role in any of the examples considered.

A state $s: MV \rightarrow D_{\mathcal I}$ is a function assigning
a domain element to each mechanism variable. Let $S_{\mathcal I}$
be the set of all states over ${\mathcal I}$. In general, whenever
the intended interpretation ${\mathcal I}$ is clear we shall tend
to omit it. The following standard logical notation will be used:
${\mathcal I},s \models \phi$ denotes that a first-order formula
$\phi$ whose variables are all in $MV$ is true in ${\mathcal I}$
at state $s$. Similarly, we let $\phi^{\mathcal I} = \{s \in
S_{\mathcal I} | {\mathcal I}, s \models \phi\}$.  Again, when the
intended interpretation is clear, we shall often simply write $s
\models \phi$.

Given interpretation ${\mathcal I}$ and an initial
state $s_{0}$, we shall interpret every mechanism $\gamma$ as a game form
of almost-perfect information $G(\gamma, s_{0}, {\mathcal I})$.
Let $\cfg$ denote the set of {\em configurations}, i.e., the set of
all pairs $\l \gamma, s \r$ where $\gamma$ is a mechanism or the empty
mechanism $\Lambda$, and $s$ is a state. We define a transition
relation $\mv{A} \subseteq \cfg \times \cfg$ for $A \subseteq Ags$ such that
$c \mv{A} c'$ states that the game can proceed from $c$ to $c'$ provided the agents
$A$ make some choice/move. In case the move does not require any agent to make
a choice, we will have $A = \emptyset$. In the standard
way (see e.g. \cite{Nielson}), we define the $\mv{A}$ relations inductively as
the smallest sets satisfying the following axioms and inference rules, the only
novelty here being the definition for $\mathtt{ch}_A$:
\[ \begin{array}{|cc|}
\hline
    \multicolumn{2}{|c|}{
    \l x:=t, s \r  \mv{\emptyset}  \l \Lambda, s^x_t \r} \\[5ex]

    \multicolumn{2}{|c|}{\l \mathtt{ch}_A(X), s \r
     \mv{A}
    \l \Lambda, s' \r \mbox{ where $s'(y)=s(y)$ for all $y \not\in X$}} \\[5ex]

    \dfrac{\l \gamma_{1}, s \r  \mv{A}  \l \Lambda, s' \r}{\l \gamma_{1}
    ; \gamma_{2} , s \r  \mv{A}  \l \gamma_{2}, s' \r} &
    \dfrac{\l \gamma_{1}, s \r  \mv{A}  \l \gamma_{1}' , s' \r}{\l
    \gamma_{1} ; \gamma_{2}, s \r  \mv{A}  \l \gamma_{1}' ; \gamma_{2}, s' \r}
    \\[5ex]

    \multicolumn{2}{|c|}{\dfrac{{\mathcal I},s \models B}{\l \mathtt{if} \: B \:
    \mathtt{then} \: \gamma_{1} \: \mathtt{else} \: \gamma_{2}, s \r
    \mv{\emptyset}  \l \gamma_{1}, s \r}} \\[5ex]
    \multicolumn{2}{|c|}{\dfrac{{\mathcal I},s \not\models B}{\l \mathtt{if} \: B \:
    \mathtt{then} \: \gamma_{1} \: \mathtt{else} \: \gamma_{2}, s \r
    \mv{\emptyset}  \l \gamma_{2}, s \r}} \\[5ex]

    \dfrac{{\mathcal I},s \not\models B}{\l \mathtt{while} \: B \:
    \mathtt{do} \: \gamma, s \r  \mv{\emptyset}  \l \Lambda, s \r} &
    \dfrac{{\mathcal I},s \models B}{\l \mathtt{while} \: B \: \mathtt{do}
    \: \gamma, s \r  \mv{\emptyset}  \l \gamma ; \mathtt{while} \: B \: \mathtt{do}
    \: \gamma, s \r}
\\[3ex]  \hline
\end{array} \]
where $s^x_t(y)=s(y)$ for $y \neq x$ and $s^x_t(x)=t^{{\mathcal I}, s}$, the
interpretation of $t$ in ${\mathcal I}$ at $s$.

Let $\cfg^*$ be the set of all finite nonempty sequences of
configurations $c_0, c_1,$ $\ldots, c_n$ such that $c_i = \l
\gamma_i, s_i \r$ and
\[ \l \gamma_0, s_0 \r \mv{A_1} \l \gamma_1, s_1 \r \mv{A_2} \ldots
\mv{A_{n}} \l \gamma_n, s_n \r ,\] and let $\cfg^*_a$ be those
sequences which end in a configuration $c_n$ for which there is
some configuration $c_{n+1}$ and set $A \subseteq Ags$ such that
$c_n \mv{A} c_{n+1}$ and $a \in A$. Infinite configuration
sequences as well as finite configuration sequences $c_0, \ldots,
c_n$ for which there is no $c_{n+1}$ and $A$ such that $c_n \mv{A}
c_{n+1}$ are called {\em terminal}, and we denote the set of
terminal sequences as $\cfg^t$.

The move relations give rise to the game tree or {\em semi-game}
$G(\gamma, s_{0}, {\mathcal I})$ which starts at the initial
position/configuration $\l \gamma, s_{0} \r$. We interpret
$\cfg^*$ as the set of (partial) histories of the game, where each
agent $a$ gets to move at the positions which are in $\cfg^*_a$.
Note that we talk of a tree, since we can think of possible loops
as infinite branches. While we shall usually refer to $G(\gamma,
s_0, {\mathcal I})$ as a game (omitting the ``semi''), note that a
semi-game lacks a link between runs/histories and preferences, for
although ${\mathcal I}$ does contain information about the
players' preferences over outcomes, the triple $G$ does not have
any mapping between histories of the game and outcomes. Such a
mapping $\widehat{o}$ will be added shortly.

A {\em strategy} for agent $a$ in semi-game $G(\gamma_0, s_{0}, {\mathcal I})$ is a
function $\sigma^{a} : \cfg^*_a \rightarrow D_{\mathcal I}$.
Given a strategy profile  $\sigma = (\sigma^1, \ldots, \sigma^n)$, i.e., a strategy
$\sigma^{a}$ for every agent $a \in Ags$, we obtain
a unique (possibly infinite) {\em run} which we denote
as $run(\sigma)$, i.e., a maximal sequence of
configurations
\[ \l \gamma_0, s_0 \r \mv{A_1} \l \gamma_1, s_1 \r \mv{A_2} \ldots \]
where $\l \gamma_0, s_0 \r$ is the initial configuration, and for
all $A_{k+1} \neq \emptyset$ we have $s_{k+1}(x_i) = \sigma^i(\l
\gamma_0, s_0 \r, \ldots, \l \gamma_k, s_k \r)$ for all $i \in A$,
and $s_{k+1}(y)=s_k(y)$ otherwise. If $run(\sigma)$ is finite, we
let $s_{\sigma}$ denote the state associated to the last
configuration of $run(\sigma)$.

\subsection*{Preferences, Predicates, and Strategic Equilibria}

Each agent has certain preferences over the various possible
outcomes of the mechanism.  Given interpretation ${\mathcal I}$
and two outcomes $o,o' {\in} D_{\mathcal I}$, agent $i$ prefers
$o$ at least as much as $o'$ whenever $o \geq_i^{\mathcal I} o'$
holds. Often, the elements in $D_{\mathcal I}$ will be elements of
some product space, so that, e.g., $(o_1, o_2) \in \mathbb{R}
\times \mathbb{R}$ will yield outcome $o_1$ for player 1 and
outcome $o_2$ for player 2, where $(o_1,o_2) \geq_i (o'_1, o'_2)$
iff $o_i \geq o'_i$.

An {\em outcome function} $\widehat{o}: \cfg^t \rightarrow
D_{\mathcal I}$ assigns an outcome to every terminal history, and
we let $\widehat{O}$ denote the the set of all outcome functions.
Given a semi-game $G(\gamma, s, {\mathcal I})$ we then obtain a
{\em game} $G(\gamma, s, {\mathcal I}, \widehat{o})$, where for
each terminal sequence of configurations $\bar{c}$ the associated
outcome is $\widehat{o}(\bar{c})$, and agent $i$ prefers
$\bar{c}_1$ to $\bar{c}_2$ iff $\widehat{o}(\bar{c}_1) \geq_i
\widehat{o}(\bar{c}_2)$. Given profile $\sigma$, we usually write
$\widehat{o}(\sigma)$ instead of $\widehat{o}(run(\sigma))$, as we
shall not be very careful about distinguishing $\sigma$ from
$run(\sigma)$.

Subgames of games will play a special role in the equilibrium
notion to be defined subsequently. A game $G'(\gamma', s',
{\mathcal I}, \widehat{o}|G')$ is a {\em subgame} of a game
$G(\gamma, s, {\mathcal I}, \widehat{o})$ iff there is a finite
sequence of configurations $ \l \gamma_0, s_0 \r \mv{A_1} \l
\gamma_1, s_1 \r \mv{A_2} \ldots \mv{A_n} \l \gamma_n, s_n \r$ for
some $n \geq 0$ such that $\l \gamma_0, s_0 \r = \l \gamma, s \r$
and $\l \gamma_n, s_n \r = \l \gamma', s' \r$. The outcome function 
$\widehat{o}|G'$ is the restriction of $\widehat{o}$ to $G'$, i.e.,
$\widehat{o}|G'(\l \gamma_n, s_n \r,\ldots,$ 
$\l \gamma_{n+k}, s_{n+k} \r) = \widehat{o}(\l \gamma_0, s_0,
\r, \ldots, \l \gamma_n, s_n \r, \ldots, \l \gamma_{n+k}, s_{n+k}
\r)$. Similarly for a strategy
profile $\sigma$ for $G$, we let $\sigma|G'$ denote its
restriction to $G'$, where $\sigma^a|G'(\l \gamma_n, s_n \r,$
$\ldots, \l \gamma_{n+k}, s_{n+k} \r) = \sigma^a(\l \gamma_0, s_0,
\r, \ldots, \l \gamma_n, s_n \r, \ldots, \l \gamma_{n+k}, s_{n+k}
\r)$.

Now that we have defined how executions of mechanisms give rise to
game trees, we can apply two well-known equilibrium notions from
game theory (see, e.g., \cite{Osborne} for a discussion of these
notions). Given a strategy profile $\sigma = (\sigma^1, \ldots,
\sigma^n)$ and a strategy $\tau^i$ for player $i$, let $(\tau^i,
\sigma^{-i})$ denote the modified strategy profile $(\sigma^1,
\ldots, \sigma^{i-1}, \tau^i, \sigma^{i+1}, \ldots, \sigma^n)$.
Furthermore, let $\sigma {\sim_{i}} \tau$ denote that the strategy
profiles $\sigma$ and $\tau$ differ at most regarding the strategy
prescribed for player $i$. Considering any game $G(\gamma,s,
{\mathcal I}, \widehat{o})$, we call a strategy profile $\sigma$ a
{\em Nash equilibrium} (NE) in $G$ iff for all agents $i$ and
strategies $\tau^i$ we have $\widehat{o}(\sigma)
\geq_{i}^{\mathcal I} \widehat{o}((\tau^{i}, \sigma^{-i}))$.
Furthermore, $\sigma$ is a {\em subgame-perfect equilibrium} (SPE)
iff for every subgame $G'$ of $G$, $\sigma|G'$ is a Nash
equilibrium in $G'$.

\bigskip

We shall usually obtain an outcome function $\widehat{o}$ from an
extended predicate, to be explained now. Given a state 
$s: MV \rightarrow D_{\mathcal
I}$ and an outcome $o \in D_{\mathcal I}$, we call $(s,o)$ an {\em
extended state}, or {\em e-state} for short. A {\em predicate} on
${\mathcal I}$ is simply a set of states $P \subseteq S_{\mathcal
I}$, and hence every FOL formula $\phi$ containing only variables
of $MV$ gives rise to a predicate $\phi^{\mathcal I}$. Similarly,
an {\em extended predicate}, or {\em e-predicate} for short, is a
set of $e$-states $P \subseteq S_{\mathcal I} \times D_{\mathcal
I}$, and every FOL formula which contains variables of $MV$ plus a
new outcome variable $x_o \not\in MV$ gives rise to an
e-predicate. We say that e-predicate $P$ is {\em functional} iff for every $s
{\in} S_{\mathcal I}$ there exists a unique $o {\in} D_{\mathcal
I}$ such that $(s,o) {\in} P$. Given two predicates (or
alternatively, two e-predicates), intersection, complementation,
etc.\ can be defined simply set-theoretically. Given a predicate
$P_{1}$ and an e-predicate $P_{2}$, however, we define $P_{1} \cap
P_{2} = \{(s,o) \in P_{2} | s \in P_{1}\}$.

Games can be obtained from extended predicates as follows: Given
semi-game $G(\gamma, s, {\mathcal I})$ and e-predicate $Q$, let
$\widehat{O}_Q$ contain all the outcome functions $\widehat{o}$ 
which assign an outcome satisfying $Q$ to every finite history, i.e.,
\[ \widehat{O}_Q = \{\widehat{o} \in \widehat{O} \; | \;
\forall run(\sigma) \in \cfg^t: \; \mbox{ if $run(\sigma)$ is
finite then } (s_{\sigma}, \widehat{o}(\sigma)) \in Q\}.\]
Note that in general,
$\widehat{O}_Q$ may be empty or contain multiple outcome
assignments. But given e-predicate $Q$ and some $\widehat{o}_Q \in
\widehat{O}_Q$, we are able to turn the semi-game $G(\gamma,s,{\mathcal I})$
into a game $G(\gamma, s, {\mathcal I}, \widehat{o}_Q)$.

%
%
\section{Mechanism Correctness}
\label{mech-des}
%
%

\subsection{Hoare Logic: From Programs to Games}
\label{ext-int}

Hoare in \cite{Hoare-logic} introduced correctness assertions of
the form $\{P\}\gamma \{Q\}$, where $\gamma$ is a program and $P$
and $Q$ are predicates. The intended interpretation of this
assertion is that in every state which satisfies $P$, any
terminating execution of program $\gamma$ ends in a state which
satisfies $Q$. In this paper, we shall extend this approach to
reason about the correctness of game-theoretic mechanisms under
subgame-perfect equilibria.

In lifting standard Hoare triples to games we generalise them in
two ways. We can view the postcondition $Q$ as
specifying the winning condition for the game, i.e., all plays of
the game ending in a state which satisfies $Q$ are a win, all
others a loss. Note that under the partial correctness reading,
infinite runs are in fact also treated as wins. Our first
generalisation consists of moving from simple win/loss situations,
represented by predicates, to general preference structures. This
is achieved by moving from predicates to e-predicates which also
specify the outcome or payoff at a state. Second and more
importantly, we move from simple claims about the existence of a
strategy profile satisfying the postcondition to more refined
claims about the existence of a strategy profile which has an
equilibrium property. This equilibrium property is generally quite
complex, and it is the complexity of this equilibrium property
which can present a challenge to compositionality, in particular
to the Hoare inference rule for composing two programs/games (see
lemma \ref{sound-comp} below).

Before defining our mechanism correctness assertions
$\{P\}\gamma\{Q\}$, it is important to point out that we are
following an {\em extensional} rather than an {\em intensional}
approach (see also \cite{Nielson}). We assume that pre- and
postconditions are predicates, i.e., semantic objects rather than
formulas of some logical language. Naturally, this means that the
calculus we present later is not fully syntactic. In the
intensional approach, however, one runs into the problem of {\em
expressiveness}, since it may happen that under a given
interpretation the logical language is not rich enough to express
all the preconditions needed. This complicates completeness proofs
considerably, due to the need for an arithmetisation of syntax
(G\"{o}delisation), etc. Furthermore, we feel that this extra
work yields more insights about the logic used for the assertion
language (usually first-order logic) than about the game theoretic
mechanisms and their equilibria, which is what we are interested
in here.

Due to its fully syntactic nature, it does seem likely that the
automated verification of mechanisms would benefit from using the
intensional approach, and we do intend to investigate this
approach in the future (see also comments in the last section).
However, note that in contrast to most computer programs whose
domain of computation contains at least the natural numbers,
mechanisms like voting procedures often use a finite domain
of computation, e.g., because there is only a small number of
possible candidates running for president. In such cases, it may
in fact be easier to do automatic verification using the
extensions of the predicates directly. Second, even if this is not
the case, the best logic to choose for automated verification may
very much depend on the class of mechanisms under consideration,
the theorem prover to be used, etc. Hence, for our present
purposes, we decide to postpone these issues since they are more 
relevant for implementation, and the extensional approach conveniently
allows us to do so.

\subsection{Mechanism Correctness and Implementation}

Assume that we are given some interpretation ${\mathcal I}$, a
mechanism $\gamma$, and e-predicates $P$ and $Q$. Then we say that
$\{P\} \gamma \{Q\}$ is {\em valid} in ${\mathcal I}$, denoted as
$ {\mathcal I} \models \{P\} \gamma \{Q\}$, iff
\begin{quote}
    for every $(s,o) \in P$, there is an outcome function
    $\widehat{o} \in \widehat{O}_Q$ and a strategy profile $\sigma$ such
    that $\sigma$ is an SPE in $G(\gamma, s, {\mathcal I},
    \widehat{o})$ and $\widehat{o}(\sigma)=o$.
\end{quote}

The notion defined indeed generalises the standard partial
correctness assertions of Hoare in the following way: Given an
arbitrary element $d \in D_{\mathcal I}$ and a predicate $P
\subseteq S_{\mathcal I}$, let $P^* = \{(s,d) | s \in P\}$. Then
given any program $\gamma \in \mbox{MPL(PRG)}$ and predicates $P$
and $Q$, the partial correctness assertion $\{P\} \gamma \{Q\}$
holds in interpretation ${\mathcal I}$ iff ${\mathcal I} \models
\{P^*\} \gamma \{Q^*\}$.

In order to link our mechanism correctness assertion to the
game-theoretic literature on mechanism design and implementation
theory \cite{Osborne,Moore:implementation,RoE}, we shall define
our version of the mechanism design problem more formally. Given a
set of possible outcomes $D_{\mathcal I}$ of the mechanism and the
set of preference profiles over $D_{\mathcal I}$, a {\em social
choice correspondence} $f$ maps a preference profile $(\geq_i)_{i
\in Ags}$ to a set of outcomes $X \subseteq D_{\mathcal I}$. The
idea is that at preference profile $(\geq_i)_{i \in Ags}$, society
or the mechanism designer wants one of the outcomes in
$f((\geq_i)_{i \in Ags})$ to be implemented or achieved. In case
$f((\geq_i)_{i \in Ags})$ is empty, society is indifferent to the
outcome actually realised. The mechanism design problem is to find
a mechanism which implements the social choice correspondence in a
non-centralised manner, i.e., no matter what the preferences of
the agents are, self-interested agents will have an incentive to
play so that the outcome intended by the designer will obtain. We
shall now see how this problem can be translated into our
mechanism correctness assertions.

For a preference profile $(\geq_i)_{i \in Ags}$ where each $\geq_i
\subseteq D_{\mathcal I} \times D_{\mathcal I}$, let ${\mathcal
I}[(\geq_i)_{i \in Ags}]$ denote the model which is obtained from
${\mathcal I}$ by
replacing the interpretation of the preference relations by the
$\geq_i$. Furthermore, for a given social choice correspondence
$f$, let $f^*(x)= \{(s,o) \in S_{\mathcal I} \times D_{\mathcal
I}| o \in f(x)\}$, and let $Q$ be any functional e-predicate. Then
we say that the pair $(\gamma,Q)$ {\em SPE-implements} a social
choice correspondence $f$ iff for all preference profiles
$(\geq_i)_{i \in Ags}$ we have
\[{\mathcal I}[(\geq_i)_{i \in Ags}] \models \{f^*((\geq_i)_{i \in
Ags})\} \gamma \{Q\}.\] To see what this statement actually
expresses, let us unpack the definition: $(\gamma, Q)$
SPE-implements social choice correspondence $f$ iff
\begin{quote}
for all preference profiles $(\geq_i)_{i \in Ags}$, for all states
$s \in S_{\mathcal I}$, and for all $o \in f((\geq_i)_{i \in
Ags})$, there is some $\widehat{o} \in \widehat{O}_Q$ and some
strategy profile $\sigma$ such that $\sigma$ is an SPE for
$G(\gamma, s, {\mathcal I}[(\geq_i)_{i \in Ags}], \widehat{o})$
and $\widehat{o}(\sigma)=o$.
\end{quote}
Note that this notion of implementation is a weak notion which
does not ask every but only some equilibrium profile to yield the
desired outcome, hence strictly speaking we are dealing with
mechanism design rather than implementation theory. In the
remainder of this section, we shall look at a few concrete
examples of mechanism design.

\subsection{Auctions}
\label{auctions}

Over the domain of natural numbers, the mechanism
\[\mathtt{ch}_{\{1,2\}}(\{x_1,x_2\})\] can represent a sealed-bid
auction where the two players simultaneously choose their bids, e.g.,
in euros, in order to obtain some desirable object, say a piano.
Since this game is atomic, the notions of SPE and NE coincide, and
hence we can phrase the existence of Nash equilibria using the
correctness notion defined earlier.

Consider the case of a second-price auction where the player
who makes the highest bid has to pay the price of the loser's bid.
We assume that our model ${\mathcal I}$ has the natural
numbers as its domain, and contains two constants $v_1$ and $v_2$
whose values denote the private valuations of the players. Instead
of representing outcomes as pairs $o=(o_1,o_2)$ we shall assume
that there are two outcome variables $o_1$ and $o_2$ which
determine the payoffs of player 1 and 2, respectively. A player's payoff 
is 0 if he fails to obtain the piano, and his valuation minus the other player's
bid if he does obtain the piano. The
preference ordering over elements of the domain is the obvious
one: $d_1 \geq_i d_2$ iff $d_1 \geq d_2$. Note that a player's preference relation
is completely determined by his valuation. 

The postcondition of
the second-price auction is the e-predicate expressed by the following formula
$\psi$:
\[ (x_1 \geq x_2 \rightarrow (o_1=v_1-x_2 \wedge o_2=0)) \wedge
    (x_1 < x_2 \rightarrow (o_1=0 \wedge o_2=v_2-x_1)) \]
It is easy to see that this postcondition expresses the payoffs of
the players in the second-price auction. Note also that the postcondition
fomalises the tie-breaking rule which assigns the object to player 1 in
case the bids are equal. Now consider the
e-predicate expressed by the following formula $\phi$:
\[ (v_1 \geq v_2 \rightarrow (o_1=v_1-v_2 \wedge o_2=0)) \wedge
    (v_1 < v_2 \rightarrow (o_1=0 \wedge o_2=v_2-v_1)) \]
We claim that ${\mathcal I} \models \{\phi^{\mathcal I}\}
\mathtt{ch}_{\{1,2\}}(\{x_1,x_2\}) \{\psi^{\mathcal I}\}$: If player
1's valuation is at least as high as player 2's valuation, then the
auction has a Nash-equilibrium in which player 2's payoff is 0 and
player 1's payoff is the difference between the valuations.  Similarly
in case player 2's valuation is higher.

To see why this is so, note that it is a well-known result in game
theory (see, e.g., \cite{Osborne}) that in a second-price
sealed-bid auction, bidding your valuation results in a Nash
equilibrium (in fact, it is even a dominant strategy). Hence, if each
player bids $x_i=v_i$, the outcomes are the ones specified by
$\phi$, and the strategies are in equilibrium. 

In fact, from the validity of $\{\phi^{\mathcal I}\}
\mathtt{ch}_{\{1,2\}}(\{x_1,x_2\}) \{\psi^{\mathcal I}\}$
we can derive
some information about the nature of the winning strategies. For suppose
w.l.o.g.\ that $v_1 \geq v_2$. Using precondition $\phi$, we know that 
$o_1 = v_1 - v_2$ and $o_2 = 0$. Now we can distinguish two cases: In the
first case, we have a Nash equilibrium (and hence also a SPE) where player 1 bids 
less than player 2, i.e., $x_1 < x_2$. Now using the postcondition $\psi$
and the fact that the outcome variables $o_1$ and $o_2$ are never changed by 
any mechanism, we know that $o_1= 0$ and $o_2 = v_2 - x_1$. Hence $x_1 = v_2 = v_1$
and $x_2 > v_2 = v_1$, i.e., the players' valuations must be the same and
player 2 must bid higher than his valuation. It is easy to check that these bids
indeed constitute a Nash equilibrium. 
In the second case, we have a Nash equilibrium with $x_1 \geq x_2$. Again using
the postcondition, $o_2= 0$ and $o_1 = v_1 - x_2$. Hence, $x_2 = v_2$ and
$x_1 \geq v_2$. Thus, player 2 bids his valuation and player 1 bids at least player
2's valuation. Again, these bid combinations all constitute Nash equilibria, and
our intended equilibrium, where each player bids his own valuation, 
is included in this second case.

In a private-value environment, a sealed-bid second-price auction is essentially
outcome equivalent with an English auction, where bidders keep increasing the price
over a number of bidding rounds until there is no more bidder who wants to 
obtain the object for a higher price.  In an English auction, bidding
slightly more than the second-highest valuation will suffice to obtain
the object.  Analogously, we can consider a sealed-bid first-price
auction where the winner has to pay his own bid rather than the
second-highest bid.  The first-price auction is essentially outcome
equivalent to the Dutch (or descending) auction, where the auctioneer
continues to lower the price of the object until a player decides to
take the object for the current price.  If the players' valuations are
not public, the safe strategy is to stop the auction just below one's
valuation, the result being that the player with the highest valuation
will obtain the object for the price of almost his valuation.

Contrary to these results, we shall show in section \ref{apps} that 
from the perspective of SPEs, the Dutch auction is also similar to
a sealed-bid second-price auction. In order to apply SPEs as a solution concept, 
we need to assume that players' preferences are public. In an auction, this
means that players know each other's valuations. In this case, however, if
$v_1 > v_2$, player 1 can wait longer before calling out to stop the
Dutch auction, he can wait until the prices reach $v_2$ or just above. 
Hence, when preferences are public, it would seem that Dutch auction
and second-price auction share a SPE. We will verify this claim in
section \ref{apps}, thereby also obtaining the precise conditions for
this equivalence.

Finally, a further remark relating auction preconditions to the notion of 
SPE-implementation.
In a second-price auction, we want to SPE-implement the social choice correspondence
$f$ which assigns to a preference profile $(v_1,v_2)$ the outcome
$(o_1,o_2)$ with $o_1=v_1 - v_2$ and $o_2=0$ in case $v_1 \geq v_2$ and
$o_2=v_2 - v_1$ and $o_1=0$ in case $v_1 < v_2$. While the precondition $\phi$
given above does capture this social choice correspondence in an intuitive sense,
note that it is not the precondition used in our definition of SPE-implementation.
This is because SPE-implementation, as we defined it, requires a correctness
claim for each preference profile separately. In contrast, our precondition $\phi$
covers all preference profiles in one precondition, since it conditions the assigned
outcomes on the relationship between the valuation constants. This formulation
leads to a much more general result and hence is usually preferable. In the next
section, we shall present an example using the notion of SPE-implementation
literally.

\subsection{Solomon's Dilemma}
\label{sol}

The biblical dilemma of Solomon (1 Kings 3:16-28) has often been
used to illustrate the basic idea of implementation theory
\cite{Osborne,Moore:implementation}. In the same spirit, we shall
use it here to illustrate our notion of SPE-implementation. The
game-theorist will get the additional benefit of seeing a
well-known example of implementation theory translated into our
framework. Solomon's dilemma is that two women have come
before him with a small child, both claiming to be the mother of
the child.

\begin{quote}
    He sent for a sword, and when it was brought, he said,
    ``Cut the living child in two and give each woman half of it.''
    The real mother, her heart full of love for her son, said to
    the king, ``Please, Your Majesty, don't kill the child! Give it
    to her!'' But the other woman said, ``Don't give it to either of
    us; go on and cut it in two.'' Then Solomon said, ``Don't kill the
    child! Give it to the first woman, she is its real mother.''
\end{quote}

The story exemplifies the need for a mechanism very well: Since
Solomon does not know who the real mother is (i.e., he does not
know the women's preferences), he cannot impose the outcome of his
choice function directly. Rather, he needs to devise a mechanism
which will provide an incentive to the women to reveal this
information to him.

To mathematically model Solomon's situation, we consider three
outcomes: $a$ (baby is given to Anne, player 1), $b$ (baby is
given to Bess, player 2), and $c$ (baby is cut in two). Solomon
has to consider two possible situations: In case Anne is the real
mother, the preference profile is given by $\theta_1$, in case
Bess is the real mother, the preference profile is $\theta_2$.
\[ \begin{array}{rccc}
\theta_1: &  a >_1 b >_1 c & \mbox{ and } & b >_2 c >_2 a \\
\theta_2: & a >_1 c >_1 b & \mbox{ and } & b >_2 a >_2 c
\end{array} \]
Solomon's problem is to find a mechanism which implements the
social choice correspondence $f$ for which $f(\theta_1)=\{a\}$ and
$f(\theta_2)=\{b\}$. In spite of Solomon's apparent cleverness, it
turns out that $f$ is not Nash-implementable 
(see \cite{Moore:implementation} for a proof). However, by slightly
modifying the problem, one can obtain an implementation
nonetheless.

Let us consider the situation where instead of quarreling about a
child, Anne and Bess argue about who is the owner of a painting.
Furthermore, we allow Solomon to impose fines on the two women,
i.e., we allow for monetary side payments. We can then think of the
possible outcomes as triples $(x,m_1,m_2)$, where $x \in
\{0,1,2\}$ denotes who obtains the painting (0 denoting that it is
cut in two), and $m_i$ denotes the fine player $i$ has to pay to
Solomon. Now suppose that the legitimate owner of the paining has
valuation $v_H$ and the other woman has valuation $v_L$, where
$v_H > v_L > 0$. Then if player $i$ does not get the painting, her
payoff is $-m_i$. If she does get the painting, her payoff will be
$v_H-m_i$ in case she is the legitimate owner, and $v_L-m_i$
otherwise. If player $i$ is the legitimate owner, these payoffs
will then induce a preference profile $\theta_i$ in the obvious
way. In this new setup, Solomon wishes to implement the social
choice rule $f$ for which $f(\theta_i)=\{(i,0,0)\}$, i.e., the
painting is given to the legitimate owner and nobody has to pay
any fines (we assume here that Solomon does not engage in dispute
resolution to make money). More precisely, Solomon is looking for
a pair $(\gamma,Q)$ which SPE-implements $f$, i.e., for which
\[{\mathcal I}[\theta_1] \models
\{o=(1,0,0)\} \gamma \{Q\} \mbox{ and } {\mathcal I}[\theta_2]
\models \{o=(2,0,0)\} \gamma \{Q\}.\]

The following mechanism $\gamma$ achieves this goal: First, Anne
is asked whether the painting is hers or not. If she says no, the
painting is given to Bess and no fines are imposed. Otherwise,
Bess is asked the same question. If Bess answers the painting is
not hers, it is given to Anne, again without imposing any fines.
Finally, in case both players have claimed to be the owner of the
painting, Anne is fined a small amount $\varepsilon > 0$ and Bess 
gets the painting but has to pay a large amount $M$ for which
$v_L < M < v_H$. The mechanism $\gamma$ can be programmed as
follows, where we take the real numbers as our domain:
\begin{tabbing}
$\mathtt{ch}_{\{1\}}(\{x_1\})$; \\
$\mathtt{if} \: x_1 > 0 \:$ \= $\mathtt{then} \: owner:=2$  \\
    \>      $\mathtt{else} \:$ \= $\mathtt{ch}_{\{2\}}(\{x_2\});$
    \\ \>\> $\mathtt{if} \: x_2 > 0 \: \mathtt{then} \: owner:=1 \: \mathtt{else} \: owner:=0$
\end{tabbing}
As for the payoff specification, let $Q$ be the e-predicate
corresponding to the following formula:
\[ \begin{array}{rl}
& (owner=1 \rightarrow o=(1,0,0)) \\ \wedge & (owner=2 \rightarrow
o=(2,0,0)) \\ \wedge & (owner=0 \rightarrow o=(2,\varepsilon,M))
\end{array} \]
Game theoretically, it is easy to verify that for preference
profile $\theta_i$, the following game form has a subgame-perfect
equilibrium yielding outcome $(i,0,0)$. We will return to this
example in section \ref{apps} and give a formal verification of
this mechanism.

\begin{picture}(100,50)
\put(80,-10){\mbox{\includegraphics{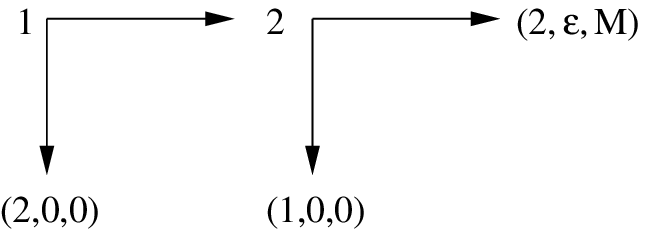}}}
\end{picture}

%
%
\section{Axiomatic Mechanism Verification}
\label{calculus}
%
%

%
\subsection{A Hoare-style Calculus}

Below we present a calculus for deriving the correctness
assertions we introduced above. Note that the calculus is a
natural generalisation of the standard Hoare calculus, where the
only addition is an axiom for the new construct $\mathtt{ch}_A$.
Given e-predicate $P$, we let $P[x/t] = \{(s,o) \in S_{\mathcal I}
\times D_{\mathcal I} | (s^x_t,o) \in P\}$.

\[ \begin{array}{|cr|}
\hline & \\
\{Q[x/t]\} x:=t \{Q\} & \mbox{(ass.)}\\[3ex]

\{wpre(ch_{A}(X), Q, {\mathcal I})\} 
ch_{A}(X) \{Q\} & \mbox{(choice)} \\[3ex]

\dfrac{\{P\} \gamma_1 \{R\} \;\;\; \{R\} \gamma_2 \{Q\}}{\{P\}
\gamma_1 ; \gamma_2 \{Q\}} & \mbox{(comp.)} \\[3ex]

\dfrac{\{P \cap B^{\mathcal I}\} \gamma_1 \{Q\} \;\;\; \{P \cap
\overline{B^{\mathcal I}}\} \gamma_2 \{Q\}}{\{P\} \mathtt{if} \: B
\: \mathtt{then} \: \gamma_{1} \: \mathtt{else} \:
\gamma_{2} \{Q\}} & \mbox{(if)} \\[5ex]

\dfrac{\{P \cap B^{\mathcal I}\} \gamma \{P\}}{\{P\}
\mathtt{while} \: B \: \mathtt{do} \: \gamma \{P \cap
\overline{B^{\mathcal
I}}\}} & \mbox{(while)} \\[3ex]

\dfrac{P \subseteq P', \; \{P'\} \gamma \{Q'\}, \; Q' \subseteq
Q}{\{P\} \gamma \{Q\}}
& \mbox{(l.c.)} \\[3ex] \hline
\end{array} \]
In the choice axiom, $wpre(\gamma, Q, {\mathcal I})$ refers to the {\em weakest precondition}
of $Q$ under $\gamma$. Given interpretation ${\mathcal I}$,
mechanism $\gamma$, and e-predicate $Q$, we define $wpre$ as
follows:
\[ \begin{array}{lcl}
    wpre(\gamma, Q, {\mathcal I}) & = & \{(s,o) \in S_{\mathcal I} \times D_{\mathcal I}\:
       | \: \exists \widehat{o} \in \widehat{O}_{Q} \exists \sigma: \;
    \sigma \mbox{ is an SPE in } \\ && \;\;\; G(\gamma, s, {\mathcal I},
    \widehat{o}) \mbox{ and } \widehat{o}(\sigma)=o\}
\end{array} \]
Note that by definition, ${\mathcal I} \models \{wpre(\gamma, Q, {\mathcal
I})\} \gamma \{Q\}$, and for every e-predicate $P$ such that
${\mathcal I} \models \{P\} \gamma \{Q\}$, we have $P \subseteq
wpre(\gamma, Q, {\mathcal I})$. Weakest preconditions will play an important
role in the completeness proof of section \ref{compl}.

Let $\Delta_{\mathcal I}$ be the smallest set of correctness
assertions $\{P\} \gamma \{Q\}$ over ${\mathcal I}$ which includes
the axioms and is closed under the inference rules above. We shall
usually write $\{P\} \gamma \{Q\} \in \Delta_{\mathcal I}$ as
${\mathcal I} \vdash \{P\} \gamma \{Q\}$. In order to gain some intuitions
regarding this calculus, the reader may wish to consult section \ref{apps}
before proceeding with the subsequent soundness and completeness results.

Before establishing soundness and completeness of the calculus presented,
some further comments regarding the choice axiom are in order.  As
mentioned, the calculus is extensional in the sense that preconditions and
postconditions are semantic rather than syntactic objects, predicates
rather than formulas of, say, first-order logic.  As a consequence, we
do not get a syntactic proof system, but rather what one might call a
compositional proof methodology.  Hence, while the precondition of the
choice axiom may seem tautological, it still suffices to reduce
reasoning about subgame-perfect equilibria in complex games to
reasoning about Nash equilibria in simple games.  Hence, while we are
still in need of a semantic argument to establish the Nash
equilibrium, it is a simpler semantic argument which applies only to
the simplest game, the atomic choice game.  As the examples in section
\ref{apps} will illustrate, this decomposition is achieved by moving
the complexity from the mechanism into the mechanism's postcondition
or payoff assignment, and it is this which the calculus allows one to
do.  In other words, the complexity is moved from the dynamic to the
static part, from the mechanism to the predicates describing pre- and
postconditions.

In verification practice, it turns out that the precondition of the
choice axiom is often rather analogous to the precondition of the
assignment axiom, where Nash equilibrium strategies are substituted
for the choice variables in the precondition.  Slightly more formally,
suppose that the postcondition $Q$ is a functional e-predicate which
simply assigns outcomes based on the choice variables, and that $Q$
only contains these choice variables and no other variables.  An
example of such a postcondition is the postcondition $\psi$ of the
second-price auction discussed in section \ref{auctions}.  Since this
postcondition depends on the state only in terms of the choice
variables, we can say that the weakest precondition of the choice
construct is simply $Q$ where each choice variable $x_{i}$ is replaced
by the Nash equilibrium strategy of player $i$ in the choice game
played in any state with payoffs given by $Q$.  In fact, this is
precisely what happened with the precondition $\phi$ of the second-price
auction where $x_{i}$ is replaced by $v_{i}$.  In general, however,
things are not quite so simple, as the analysis of the Dutch auction
in section \ref{apps} will illustrate.

\subsection{Soundness}

The following lemma presents the first of the two most difficult
cases of the subsequent soundness result. It guarantees that
equilibria of subgames can be composed into equilibria of the
supergame.

\begin{lemma}[Composition] If we have both ${\mathcal I} \models
\{P\} \gamma_1 \{R\}$ and ${\mathcal I} \models \{R\} \gamma_2
\{Q\}$ then ${\mathcal I} \models \{P\} \gamma_1 ; \gamma_2
\{Q\}$. \label{sound-comp}
\end{lemma}
\begin{proof}
    Let $(s,o) \in P$, and consider $G(\gamma_1 ;
    \gamma_2, s, {\mathcal I})$. By our first
    assumption, there is an outcome function $\widehat{o}_1 \in
    \widehat{O}_{R}$ and a strategy profile $\sigma_1$ such that $\sigma_1$
    is an SPE in $G_1(\gamma_1, s, {\mathcal I}, \widehat{o}_1)$ and
    $\widehat{o}_1(\sigma_1)=o$.

    Now for every finite run $\tau_1$ of $G_1$ ending in some terminal
    state $t$ with $\widehat{o}_1(\tau_1)=o_t$, since $(t,o_t) \in
    R$, we know by our second assumption that there is some
    outcome function $\widehat{o}_t \in \widehat{O}_{Q}$ and some strategy
    profile $\sigma_t$ such that $\sigma_t$ is an SPE in
    $G_t(\gamma_2, t, {\mathcal I}, \widehat{o}_t)$ and
    $\widehat{o}_t(\sigma_t)=o_t$. Taken together, $\sigma_1$ and
    the $\sigma_t$ induce a strategy profile $\sigma$ for $G$, and
    similarly $\widehat{o}_1$ (for the infinite runs of $G_1$) and the
    $\widehat{o}_t$ induce an outcome function $\widehat{o} \in
    \widehat{O}_Q$ for $G$. Hence, it remains to show that $\sigma$ 
    is an SPE and that $\widehat{o}(\sigma)=o$.

    First, it is easily seen that $\widehat{o}(\sigma)=o$, for
    $\widehat{o}_1(\sigma_1)=o$, and so in case $\sigma_1$ is finite,
    $(s_{\sigma_1},o) \in R$, from which by definition it follows that
    $\widehat{o}(\sigma)=o$. Second, we need to show that $\sigma$
    is an SPE in $G(\gamma_1 ; \gamma_2, s, {\mathcal I},
    \widehat{o})$. So consider any subgame $G'(\pi, t, {\mathcal I})$ of $G$. In
    the easy case, $G'$ will be a subgame of some $G_{t'}$, where $t'$
    is a terminal state in $G_1$, for in this case,
    our second assumption immediately guarantees the equilibrium property.
    In the more complicated case, $G'$ lies partly in $G_1$.  For simplicity,
    we shall for the rest of this argument assume that $\sigma = \sigma_1 {\cdot} \sigma_2$
    refers to its restriction to $G'$.  So consider any strategy profile $\tau_1
    {\cdot} \tau_2$ for $G'$ such that $\sigma = \sigma_1 {\cdot}
    \sigma_2 \sim_i \tau_1 {\cdot} \tau_2 = \tau$, where $\sigma_1$ and $\tau_1$ both
    yield finite runs.  Suppose further
    that $\widehat{o}(\sigma)=o_0$ and $\widehat{o}(\tau)=o_2$, as
    depicted below.

\begin{picture}(200,150)
\put(70,10){\mbox{\includegraphics{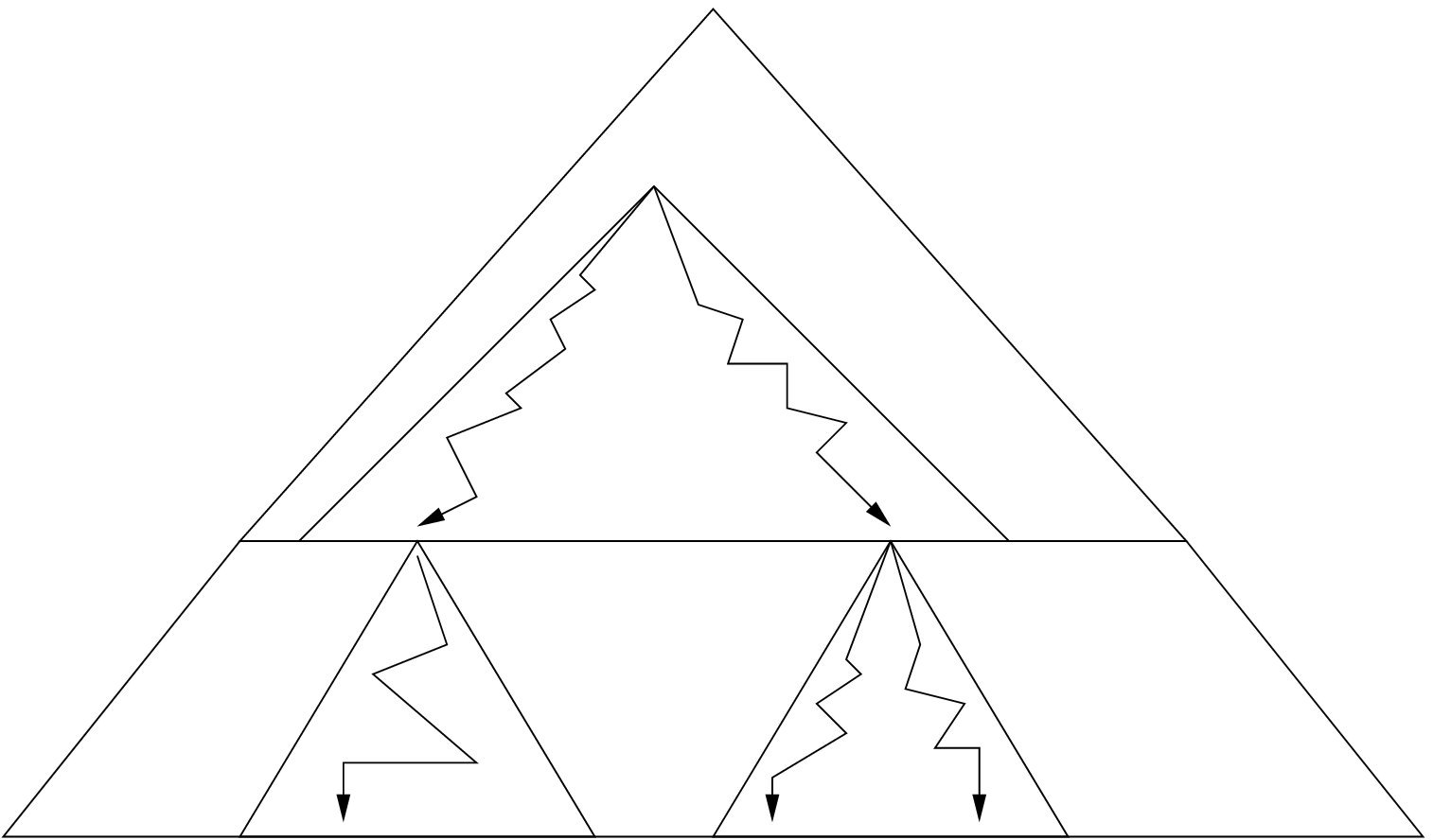}}}
\put(168,112){$t$}
\put(56,90){$\gamma_1$}
\put(56,30){$\gamma_2$}
\put(151,75){$\sigma_1$}
\put(181,75){$\tau_1$}
\put(146,61){$o_0$}
\put(186,61){$o_1$}
\put(166,61){$\geq_i$}

\put(127,15){$\sigma_2$}
\put(192,15){$\sigma_2$}
\put(207,15){$\tau_2$}

\put(120,3){$o_0$}
\put(186,3){$o_1$}
\put(202,3){$\geq_i$}
\put(217,3){$o_2$}
\end{picture}

\noindent Now supposing that $\widehat{o}(\tau_1 {\cdot}
\sigma_2)=o_1$, we know by definition of $\sigma$ that $o_1 \geq_i
o_2$, and that $\widehat{o}_1(\tau_1)=o_1$. Furthermore, since
$\sigma_1$ was an SPE in $G_1$, we know also that
$\widehat{o}_1(\sigma_1) \geq_i o_1$. Since $o_0 =
\widehat{o}(\sigma)=\widehat{o}_1(\sigma_1)$, we can conclude by
transitivity that $o_0 \geq_i o_2$.

Finally, note that the case where either $\sigma_1$ or $\tau_1$ or
both are infinite can be treated by a simplification of the above
argument.
\end{proof}

The following lemma isolates the arguments needed to prove the
soundness of the inference rule for iteration. Our assumption that
our model ${\mathcal I}$ contains a uniformly worst element is
needed here.

\begin{lemma}
If ${\mathcal I} \models \{P \cap B^{\mathcal I} \} \gamma \{P\}$
then ${\mathcal I} \models \{P\} \mathtt{while} \: B \:
\mathtt{do} \: \gamma \{P \cap \overline{B^{\mathcal I}}\}$.
\label{sound-iteration}
\end{lemma}
\begin{proof}
Roughly speaking, the proof is an iterated application of the
preceding composition lemma, but a few subtleties have to be dealt
with, in particular the possibility of newly arising infinite
runs.

Suppose that $(s,o) \in P$. In order to define a strategy $\sigma$
and outcome function $\widehat{o}$ for $G(\mathtt{while} \: B \:
\mathtt{do} \: \gamma, s, {\mathcal I})$, we shall inductively
define strategy profile $\sigma_n$ and outcome function
$\widehat{o}_n$ for game $G_n$ which consists of the first $n$
iterations of game $G$. Game $G_0$ simply consists of
configuration $(\Lambda, s)$, strategy profile $\sigma_0$ consists
of doing nothing, and as an outcome function we take
$\widehat{o}_0((\Lambda,s))=o$. Note that $\widehat{o}_0 \in
\widehat{O}_P$.

For the inductive step, define $G_{n+1}$ as $G_n$ where for every
terminal state $(t,o_t) \in P \cap B^{\mathcal I}$ in $G_n$ we
concatenate $G_t(\gamma,t,{\mathcal I})$ to $t$. By our
assumption, for each such terminal state, we have an outcome
function $\widehat{o}_t$ and a SPE strategy profile $\sigma_t$, and we
define $\sigma_{n+1}$ and $\widehat{o}_{n+1}$ in the natural way,
by extending $\sigma_n$ and $\widehat{o}_n$ to $G_{n+1}$ using the
$\widehat{o}_t$ and $\sigma_t$.

Now with slight abuse of notation, we can define strategy
profile $\sigma$ and outcome function $\widehat{o}$ for $G$ as
follows: We take $\sigma = \bigcup_i \sigma_i$, i.e., we simply
take the profile generated by the $\sigma_i$. Similarly, we define
$\widehat{o} = \bigcup_i \widehat{o}_i$, i.e., every run $\tau$ of
$G$ which is part of some $G_i$ is evaluated according to
$\widehat{o}_i$. Furthermore, there may be new infinite runs in
$G$ which are not part of any $G_i$, but are instead generated by
an infinite number of plays of $\gamma$ itself. Given such an
infinite run $\tau$, we define $\widehat{o}(\tau)=o_c$ in case
there is some $j$ such that for all $k \geq j$ we have
$\widehat{o}_k(\tau | G_k)=o_c$; otherwise, we let
$\widehat{o}(\tau)=-\infty$. Thus, for infinite runs which
converge on a certain outcome $o_c$, we assign $o_c$ to the run,
and otherwise simply the uniformly worst outcome. Note that
$\widehat{o} \in \widehat{O}_{P \cap \overline{B^{\mathcal I}}}$.

Observe first that $\widehat{o}(\sigma)=o$. For we have
$\widehat{o}_1(\sigma_1)=o$,
$\widehat{o}_2(\sigma_2)=\widehat{o}_1(\sigma_1)=o$, etc., and so
in case $\sigma$ is finite, there is some maximal $k$ such that
$\widehat{o}(\sigma)=\widehat{o}_k(\sigma_k)=o$. In case $\sigma$
is infinite, we have a constant and hence converging sequence of
outcomes consisting of $o$ only.

Hence, all we need to show is that $\sigma$ is an SPE in
$G(\mathtt{while} \: B \: \mathtt{do} \: \gamma, s, {\mathcal I},
\widehat{o})$. So consider any subgame $G'$ of $G$ and a strategy
$\tau \sim_i \sigma$ such that $\widehat{o}(\sigma)=o_0$ and
$\widehat{o}(\tau)=o_2$. Now the reasoning can proceed along the
lines of the composition lemma and the figure given there: In case
$\tau$ yields a run which lies in $G_k$, we can show by induction
on $k$ that $o_0 \geq_i o_2$, each step involving the reasoning
carried out in the composition lemma. On the other hand, in case
$\tau$ is an infinite run generated by infinitely many
$\gamma$-repetitions, we need to distinguish two cases: In the
easy case where $\widehat{o}(\tau)=-\infty$, the result is
obvious. In the more complicated case, $\widehat{o}(\tau)=o_c$ due
to a sequence of outcomes which converges on $o_c$. Suppose $k$ is
the smallest number for which $\widehat{o}_k(\tau | G_k)=o_c$.
Then again we can apply the reasoning of the composition lemma $k$
times to show that $o_0 \geq_i o_c$.
\end{proof}

\begin{theorem}[Soundness]
If ${\mathcal I} \vdash \{P\} \gamma \{Q\}$ then ${\mathcal I}
\models \{P\} \gamma \{Q\}$.
\end{theorem}
\begin{proof}
    The proof is by induction on the length of the derivation,
so we start with showing the validity of the axioms.  The soundness of
the $\mathtt{ch}_{A}$ axiom follows by definition.

For ${\mathcal I} \vdash \{Q[x/t]\} x:=t \{Q\}$, suppose that
$(s,o) \in Q[x/t]$.  We know that all runs in $G(x:=t,s,{\mathcal
I})$ are finite.  Since no choices need to be made in
$G(x:=t,s,{\mathcal I})$, the one existing strategy profile
$\sigma$ is trivially an equilibrium in $G(x:=t,s,{\mathcal
I},\widehat{o})$ for any outcome function $\widehat{o} \in
\widehat{O}_{Q}$, and in particular for the outcome function
$\widehat{o}$ which assigns $o$ to $\sigma$. Note that since
$(s,o) \in Q[x/t]$, $(s^x_t,o) \in Q$, and hence $\widehat{o} \in
\widehat{O}_Q$.

Turning to the inference rules, note that the case of composition
is treated in lemma \ref{sound-comp}, and the logical consequence
rule is an easy consequence of the semantic definition of
${\mathcal I} \models \{P\} \gamma \{Q\}$. For conditional
branching, the conclusion follows directly from the two premises,
given that $G(\mathtt{if} \: B \: \mathtt{then} \: \gamma_{1} \:
\mathtt{else} \: \gamma_{2}, s, {\mathcal I})$ is either
$G_{1}(\gamma_{1}, s, {\mathcal I})$ or $G_{2}(\gamma_{2}, s,
{\mathcal I})$.  Finally, lemma \ref{sound-iteration} takes care
of iteration.
\end{proof}

Note that the soundness result also holds for Nash equilibria: If
in the definition of ${\mathcal I} \models \{P\} \gamma \{Q\}$ we
replace SPE by NE, the above soundness result can still be proved.
This is as it should be, since every subgame-perfect equilibrium
is also a Nash equilibrium.

\subsection{Completeness}
\label{compl}

Like in the completeness proof for the standard Hoare calculus,
the notion of a weakest precondition plays an important role for
our calculus as well. The following lemma contains the essential 
argument for the completeness result.

\begin{lemma}[Decomposition]
If ${\mathcal I} \models \{P\} \gamma_1;\gamma_2 \{Q\}$, then for
some $R$ we have ${\mathcal I} \models \{P\} \gamma_1 \{R\}$ and
${\mathcal I} \models \{R\} \gamma_2 \{Q\}$.
\end{lemma}
\begin{proof}
Our assumption is ${\mathcal I} \models \{P\}
\gamma_{1};\gamma_{2} \{Q\}$. Let $R=wpre(\gamma_2, Q, {\mathcal
I})$, then all we need to show is that ${\mathcal I} \models \{P\}
\gamma_1 \{R\}$. So supposing that $(s,o) \in P$, we need to
provide an outcome function $\widehat{o}_1 \in \widehat{O}_R$ and
a strategy profile $\sigma_1$ such that $\sigma_1$ is an SPE in
$G_1(\gamma_1, s, {\mathcal I}, \widehat{o}_1)$ and
$\widehat{o}_1(\sigma_1)=o$.

Consider the outcome function $\widehat{o} \in \widehat{O}_Q$ and
the strategy profile $\sigma$ for $G(\gamma_1;\gamma_2, \linebreak
s, {\mathcal I})$ provided by our assumption. We let $\sigma_1 =
\sigma | G_1$. As for the definition of $\widehat{o}_1$, for every
infinite run $\tau$ of $G_1$ we let
$\widehat{o}_1(\tau)=\widehat{o}(\tau)$. If on the other hand
$\tau$ is finite, we define $\widehat{o}_1(\tau)=\widehat{o}(\tau
{\cdot} \sigma_{\tau})$, where $\sigma_{\tau} = \sigma |
G_{\tau}$. By our assumption, we have
$\widehat{o}_1(\sigma_1)=\widehat{o}(\sigma)=o$. Furthermore,
since $(s_{\tau},\widehat{o}(\tau {\cdot} \sigma_{\tau})) \in R$,
$\widehat{o}_1 \in \widehat{O}_R$.

Hence, all we need to show is that $\sigma_1$ is an SPE in
$G_1(\gamma_1, s, {\mathcal I}, \widehat{o}_1)$. So consider any
subgame $G'_1=(\pi, t, {\mathcal I}, \widehat{o}_1)$ of $G_1$, and
a strategy profile $\tau_1 \sim_i \sigma_1$, where we take
$\widehat{o}_1(\sigma_1)=o_0$ and $\widehat{o}_1(\tau_1)=o_1$.
Assume first that both $\sigma_1$ and $\tau_1$ are finite.
Considering $G'=(\pi;\gamma_2, t, {\mathcal I}, \widehat{o})$, we
know that there is a profile $\sigma_2$ (derived from $\sigma$)
such that $\sigma_1 {\cdot} \sigma_2$ is an SPE in $G'$ and
$\sigma_1 {\cdot} \sigma_2 \sim_i \tau_1 {\cdot} \sigma_2$. The
situation is depicted below. 

\begin{picture}(200,150)
\put(70,10){\mbox{\includegraphics{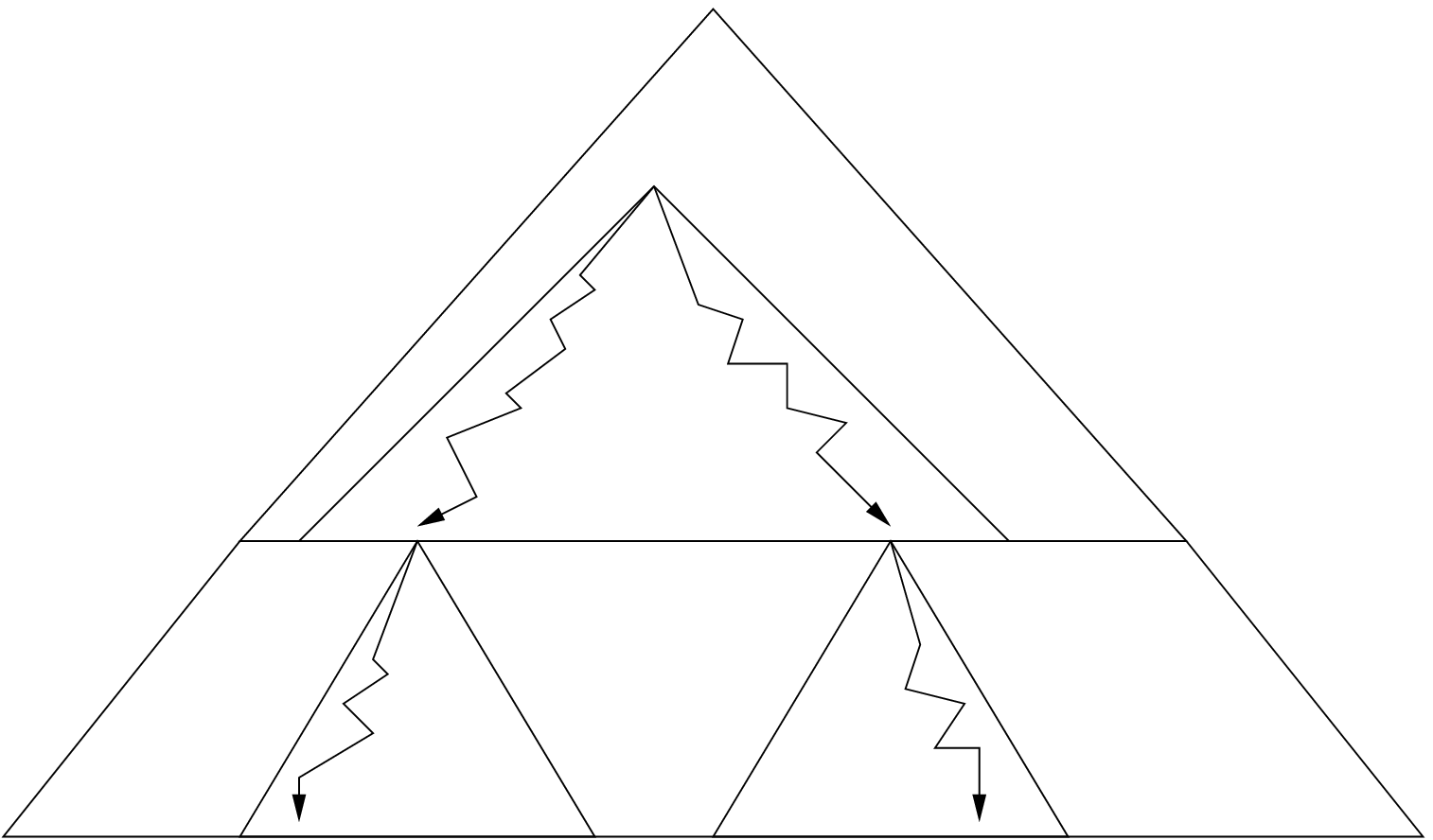}}}
\put(168,112){$t$}
\put(56,90){$\gamma_1$}
\put(56,30){$\gamma_2$}
\put(151,75){$\sigma_1$}
\put(181,75){$\tau_1$}
\put(146,61){$o_0$}
\put(186,61){$o_1$}
\put(166,61){$\geq_i$}

\put(119,15){$\sigma_2$}

\put(207,15){$\sigma_2$}

\put(112,3){$o_0$}

\put(217,3){$o_1$}
\end{picture}

By definition, we know that $\widehat{o}_1(\sigma_1) =
\widehat{o}(\sigma_1 {\cdot} \sigma_2)=o_0$ and
$\widehat{o}_1(\tau_1) = \widehat{o}(\tau_1 {\cdot}
\sigma_2)=o_1$, and hence we must have $o_0 \geq_i o_1$.

Note that in case either $\sigma_1$ or $\tau_1$ or both are
infinite, a simplified version of the above argument can be
applied.
\end{proof}

The above lemma is what distinguishes subgame-perfect equilibria
from Nash equilibria, since only the former can be decomposed in
the way shown by the decomposition lemma. For Nash equilibria, the
above lemma fails: when defining $\widehat{o}_1$ in the above
proof, we cannot be sure that $\widehat{o}_1 \in \widehat{O}_R$,
since a subprofile of an equilibrium profile may itself not be an
equilibrium profile. Consequently, also the following completeness
result does not hold for Nash equilibria.

\begin{theorem}[Completeness]
If ${\mathcal I} \models \{P\}\gamma\{Q\}$ then
${\mathcal I} \vdash \{P\}\gamma\{Q\}$.
\end{theorem}
\begin{proof}
The proof proceeds by induction on the structure of $\gamma$. For
$x:=t$, note that for any state $s$, the game $G(x:=t, s,
{\mathcal I})$ contains only a single finite run ending in state
$s^x_t$. Observe that $P \subseteq Q[x/t]$: if $(s,o) \in P$,
every run terminates in state $(s^x_t,o) \in Q$, and hence $(s,o)
\in Q[x/t]$. Applying the logical consequence rule to the
assignment axiom, we then obtain ${\mathcal I} \vdash \{P\} x:=t
\{Q\}$.

For $\mathtt{ch}_A$, we use the axiom and the logical consequence rule,
and for $\gamma_1 ; \gamma_2$, we can appeal to the decomposition
lemma, induction hypothesis, and the composition rule.  The case of
$\mathtt{if} \: B \: \mathtt{then} \: \gamma_{1} \: \mathtt{else} \:
\gamma_{2}$ is straight-forward, so we only need to deal with the
while-loop.

For iteration, suppose that ${\mathcal I} \models \{P\}
\mathtt{while} \: B \: \mathtt{do} \: \gamma \{Q\}$.  Similarly,
to the proof of the decomposition lemma, we let $R =
wpre(\mathtt{while} \: B \: \mathtt{do} \: \gamma, Q, {\mathcal
I})$. First, we shall establish that ${\mathcal I} \models \{R
\cap B^{\mathcal I}\} \gamma \{R\}$.  By definition, we have
${\mathcal I} \models \{R\} \mathtt{while} \: B \: \mathtt{do} \:
\gamma \{Q\}$.  From this, ${\mathcal I} \models \{R \cap
B^{\mathcal I}\} \gamma ; \mathtt{while} \: B \: \mathtt{do} \:
\gamma \{Q\}$ is easily seen to follow.  Now we can apply the
decomposition lemma: Since the $R$ provided by the proof of the
decomposition lemma is precisely the one we defined above, we can
conclude that ${\mathcal I} \models \{R \cap B^{\mathcal I}\}
\gamma \{R\}$.

Now using the induction hypothesis and applying the while-rule, we obtain
\[ {\mathcal I} \vdash \{R\} \mathtt{while} \: B \:
 \mathtt{do} \: \gamma \{R \cap \overline{B^{\mathcal I}}\}.\]
Since $P \subseteq R$ and $R \cap \overline{B^{\mathcal I}}
\subseteq Q$, we can apply the logical consequence rule to derive
${\mathcal I} \vdash \{P\} \mathtt{while} \: B \: \mathtt{do} \:
\gamma \{Q\}$.
\end{proof}

%
%
\section{Applying the Calculus - Some Examples}
\label{apps}
%
%

%
\subsection{Solomon's Dilemma}

Consider again Solomon's 2-stage mechanism given in 
section \ref{sol}, where we will replace the variable
$owner$ by $w$ to save space. We will show one of the two required
correctness claims, namely that ${\mathcal I}[\theta_1] \vdash$
\begin{tabbing}
$\{o=(1,0,0)\}$ \\
$\mathtt{ch}_{\{1\}}(\{x_1\})$; \\
$\mathtt{if} \: x_1 > 0 \:$ \= $\mathtt{then} \: w:=2$  \\
    \>      $\mathtt{else} \:$ \= $\mathtt{ch}_{\{2\}}(\{x_2\});$
    \\ \>\> $\mathtt{if} \: x_2 > 0 \: \mathtt{then} \: w:=1 \: 
          \mathtt{else} \: w:=0$ \\
$\{(w=1 \rightarrow o=(1,0,0)) \wedge  (w=2 \rightarrow
o=(2,0,0)) \wedge (w=0 \rightarrow o=(2,\varepsilon,M))\}$,
\end{tabbing}
corresponding to the situation where player 1 is the real owner of
the painting.  Note that for ease of notation we are now simply
representing (extended) predicates by formulas in first-order logic.

Denoting the postcondition by $Q_0$, we have ${\mathcal I}[\theta_1] \vdash
\{o=(2,\varepsilon,M)\} w:=0 \{Q_0\}$ and ${\mathcal I}[\theta_1] \vdash
\{o=(1,0,0)\} w:=1 \{Q_0\}$ using the assignment axiom. Hence, by
the if-rule we have ${\mathcal I}[\theta_1] \vdash$ 
\begin{tabbing}
$\{(x_2 > 0 \rightarrow o=(1,0,0)) \wedge (x_2 \leq 0 \rightarrow o=(2,\varepsilon,M))\}$  
\\ $\mathtt{if} \: x_2 > 0 \: \mathtt{then} \: w:=1 \: \mathtt{else} \: w:=0$ \\
$\{Q_0\}.$
\end{tabbing}
Denote the new precondition by $Q_1$. Since in $\theta_1$, we have $(1,0,0)
>_2 (2,\varepsilon,M)$, we know that when choosing a value for $x_2$, player
2 will choose the outcome $(1,0,0)$, and hence we have ${\mathcal I}[\theta_1]
\vdash \{o=(1,0,0)\} \mathtt{ch}_{\{2\}}(\{x_2\}) \{Q_1\}$.  On the
other hand, we know by the assignment rule that ${\mathcal
I}[\theta_1] \vdash \{o=(2,0,0)\} w:=2 \{Q_0\}$.  Hence, using the
if-rule and composition, we have ${\mathcal I}[\theta_1] \vdash$
\begin{tabbing}
$\{(x_1 > 0 \rightarrow o=(2,0,0)) \wedge (x_1 \leq 0 \rightarrow o=(1,0,0))\}$\\  
$\mathtt{if} \: x_1 > 0 \:$ \= $\mathtt{then} \: w:=2$  \\
    \>      $\mathtt{else} \:$ \= $\mathtt{ch}_{\{2\}}(\{x_2\});$
    \\ \>\> $\mathtt{if} \: x_2 > 0 \: \mathtt{then} \: w:=1 \: 
          \mathtt{else} \: w:=0$ \\
$\{Q_0\},$
\end{tabbing}
where we denote the new precondition by $Q_2$.
Finally, since $(1,0,0) >_1 (2,0,0)$, player 1 will choose $(1,0,0)$ in an
equilibrium,  and so we have 
${\mathcal I}[\theta_1] \vdash \{o=(1,0,0)\} \mathtt{ch}_{\{1\}}(\{x_1\}) \{Q_2\}$.
Using the composition rule, we have thereby succeeded in verifying the original claim,
that the 2-stage mechanism does indeed provide an SPE-implementation solving
Solomon's (modified) dilemma.

\subsection{Auctions}

\subsubsection*{Second-Price Sealed-Bid Auction}
    
We have already presented the sealed-bid second-price auction in section 
\ref{auctions}.
We argued that in the relevant model ${\mathcal I}$ where two players
have private valuations represented by the constants $v_{1}$ and $v_{2}$, we
have ${\mathcal I} \models$
\begin{tabbing}
$\{(v_1 \geq v_2 \rightarrow (o_1=v_1-v_2 \wedge o_2=0)) \wedge
    (v_1 < v_2 \rightarrow (o_1=0 \wedge o_2=v_2-v_1))\}$ \\
$\mathtt{ch}_{\{1,2\}}(\{x_1,x_2\})$ \\
$\{(x_1 \geq x_2 \rightarrow (o_1=v_1-x_2 \wedge o_2=0)) \wedge
    (x_1 < x_2 \rightarrow (o_1=0 \wedge o_2=v_2-x_1))\}$,
\end{tabbing}
due to the fact that we obtain a Nash equilibrium if each player bids
his valuation, i.e.\ $x_{i} = v_{i}$. We abbreviate the given
precondition with $P$ and the postcondition with $R$.  Note that $P$
is not the weakest precondition of
$G(\mathtt{ch}_{\{1,2\}}(\{x_1,x_2\}), R, {\mathcal I})$, and hence
${\mathcal I} \vdash \{P\} \mathtt{ch}_{\{1,2\}}(\{x_1,x_2\}) \{R\}$
is not an axiom.  This is because there are equilibria other than the
one mentioned.  For example, suppose that $v_{1} \geq v_{2}$.  Then if
$v_{2} \leq x_{1} = x_{2} \leq v_{1}$, we also have a Nash
equilibrium.  Hence, for $v_{2} \leq k \leq v_{1}$, we can also
consider the following precondition $P_{k}$ \[ (v_1 \geq v_2
\rightarrow (o_1=v_1-k \wedge o_2=0)) \wedge (v_1 < v_2 \rightarrow
(o_1=0 \wedge o_2=v_2-k)) \] for which we also have ${\mathcal I}
\models \{P_{k}\} \mathtt{ch}_{\{1,2\}}(\{x_1,x_2\}) \{R\}$. 
Consequently, $P_{k} \vee P$ is weaker than $P$ for $k \neq v_{2}$,
and hence ${\mathcal I} \vdash \{P\}
\mathtt{ch}_{\{1,2\}}(\{x_1,x_2\}) \{R\}$ is indeed not an axiom. 
Still, it can be easily obtained from the choice axiom using the
logical consequence rule.

\subsubsection*{Dutch Auction}

We shall now illustrate the calculus in action for verifying the more complex
Dutch auction which involves a while loop.  In fact, we shall
illustrate that the Dutch auction is equivalent to the preceding
sealed-bid second-price auction in the very weak sense that the Dutch
auction has the same subgame-perfect equilibrium as the sealed-bid
second-price auction, where the player with the higher valuation
receives the object, paying the price of the other player's valuation. 
More formally, we shall show that both implement the same social
choice correspondence defined in section \ref{auctions}, under certain conditions.

As mentioned in section \ref{auctions}, in a Dutch auction, the auctioneer
continues to lower the price of an object until a player decides to
take the object for the current price.  Over the domain of natural
numbers, the Dutch auction is captured by the following mechanism
$\alpha$:
\begin{tabbing}
$p:=init$; \\ $w:=0$; \\
$\mathtt{while}$ \= $\: p > 0 \wedge w=0 \: \mathtt{do}$ \\ \>
$\mathtt{ch}_{\{1,2\}}(\{x_1,x_2\})$; \\
    \> $\mathtt{if} \: x_1 > 0 \:$ \= $\mathtt{then} \: w:=1$ \\
    \>\> $\mathtt{else} \: \mathtt{if} \: x_2 > 0 \:$ \= $\mathtt{then}
    \: w:=2$ \\
    \>\>\> $\mathtt{else} \: p:=p-1$
\end{tabbing}
Variable $w$ keeps track of the winner, $p$ keeps track of the current
price, and is initialised to some value $init$.  For each offer, both
players can choose a nonnegative number signaling their desire to buy
the object for price $p$.  As the algorithm is written down here, in
case both players want to buy the object, player 1 gets it.  Note that
it is also subtleties like these which provide an argument for
formally specifying and verifying mechanisms.  The following
postcondition $Q$ naturally assigns payoffs at the end of the Dutch auction:
\[ \begin{array}{rl}
& (w=1 \rightarrow (o_1=v_1-p \wedge o_2=0)) \\ \wedge & (w=2
\rightarrow (o_1=0 \wedge o_2=v_2-p)) \\ \wedge & (w=0 \rightarrow
(o_1=0 \wedge o_2=0))
\end{array}    \]
Our goal will be to show that ${\mathcal I} \vdash \{P\} \alpha
\{Q\}$, i.e., just like the sealed-bid
auction $(\mathtt{ch}_{\{1,2\}}(\{x_1,x_2\}),R)$ SPE-implements our desired
social choice correspondence, so does $(\alpha,Q)$.

As in standard program verification, the art of proving the
correctness of a while-loop lies in finding an invariant which remains
true at the beginning of every loop execution.  Consider the following
invariant $Inv$:
\[ \begin{array}{l}
v_{1} \geq v_{2} > 0 \wedge p \geq v_{2} \wedge w \in \{0,1,2\} \\ \wedge \:
(w=1  \rightarrow (o_1=v_1-p \wedge o_2=0)) \\ \wedge  \: (w=2 
\rightarrow (o_1=0 \wedge o_2=v_2-p)) \\ \wedge \:
(w=0 \rightarrow (o_{1}=v_{1} - v_{2} \wedge o_{2} = 0))
\end{array}    \]

Note that in order to simplify the exposition we have restricted ourselves
to the case where $v_1 \geq v_2$, but this restriction is in no way essential.
The invariant is similar to the desired postcondition $Q$, the main
difference lies in the situation where there is no winner. In that case, our
desired outcome will be the SPE of the remaining subgame, the outcome
designated by our social choice function, $o_{1} = v_{1} - v_{2}$ and
$o_{2} = 0$. Besides these winning conditions, we state the range of variable 
$w$ as well as two conditions for $v_2$. First, $v_2$ must never be greater than 
the current price, for our equilibrium strategies force us to exit the loop
at $v_2$. If, e.g., the auction started with a price below $v_2$, player 1 could
immediately take the object and thereby receive a payoff higher than $v_1-v_2$.
Second, $v_2$ must be strictly greater than 0, for otherwise, it would be optimal 
for player 1 to take the object in the last round, where the price $p=1$, and hence
obtaining a payoff lower than $v_1-v_2$. Note that the need for these additional 
constraints was discovered in the verification process and hence the ``discovery''
of these crucial side conditions should be regarded as a result of the 
verification effort.

We will now proceed to show that $Inv$ is indeed an
invariant, i.e., that ${\mathcal I} \vdash$
\begin{tabbing}
 $\{Inv \wedge p>0 \wedge w=0\}$ \\
$\mathtt{ch}_{\{1,2\}}(\{x_1,x_2\})$; \\
    $\mathtt{if} \: x_1 > 0 \:$ \= $\mathtt{then} \: w:=1$ \\
    \> $\mathtt{else} \: \mathtt{if} \: x_2 > 0 \:$ \= $\mathtt{then}
    \: w:=2$ \\
    \>\> $\mathtt{else} \: p:=p-1$ \\
$\{Inv\}$
\end{tabbing}
Note that in fact, $p>0$ is already implied by $Inv$ which means that
if $Inv$ is indeed an invariant, the auction can never terminate due
to the price having reached 0.  Hence, for the purposes of verifying the
desired equilibrium, the condition $p>0$ is redundant in the guard
condition of the while-loop.

To begin with, applying the assignment rule and the if-rule, it is
easy to check that  ${\mathcal I} \vdash$
\begin{tabbing}
$\{ v_{1} \geq v_{2} > 0 \wedge p \geq v_{2} \wedge
(x_{1} > 0 \rightarrow (o_1=v_1-p \wedge o_2=0))$ \\ $\wedge \: ((x_{1}=0
\wedge x_{2} > 0) \rightarrow (o_1=0 \wedge o_2=v_2-p))$ \\ $\wedge \:
((x_{1}=0 \wedge x_{2} = 0) \rightarrow Inv[p/p-1])\}$ \\
    $\mathtt{if} \: x_1 > 0 \:$ \= $\mathtt{then} \: w:=1$ \\
    \> $\mathtt{else} \: \mathtt{if} \: x_2 > 0 \:$ \= $\mathtt{then}
    \: w:=2$ \\
    \>\> $\mathtt{else} \: p:=p-1$ \\
$\{Inv\},$
\end{tabbing}
where $Inv[p/p-1]$ results from substituting $p-1$ for $p$ in
$Inv$.  Denote the new precondition as $Inv_{2}$.  Now we claim that
${\mathcal I} \vdash$ 
\begin{tabbing}
    $\{v_{1} \geq v_{2} > 0 \wedge p \geq v_{2} \wedge w=0$ \= $\wedge \:
    (p \leq v_{2} \rightarrow (o_{1}=v_{1} - p \wedge o_{2} = 0))$\\
    \> $\wedge \: (p>v_{2} \rightarrow (o_{1}=v_{1} - v_{2} \wedge o_{2} =
    0))\}$ \\
    $\mathtt{ch}_{\{1,2\}}(\{x_1,x_2\})$ \\
    $\{Inv_{2}\}$
\end{tabbing}
Assume that $v_{1} \geq v_{2} > 0$, and consider a state $s$ where $p \geq
v_{2}$ and $w=0$.  We distinguish two cases.  First, if $p \leq v_{2}$
(i.e., $p=v_{2}$), both players asking for the object, i.e., $x_{1} >
0$ and $x_{2} > 0$, constitutes a Nash equilibrium in the game with
payoffs according to $Inv_{2}$, with payoffs $o_{1}= v_{1}-p$ and
$o_{2} = 0$.  Second, suppose that $p > v_{2}$.  In this case, both
players declining the object, i.e., $x_{1} = x_{2} = 0$, constitutes a
Nash equilibrium.  Player 2 should not ask for it since the price
exceeds his valuation, and player 1 should not ask for it since the
price will be lower in the next round; formally, declining the object
yields $o_{1} = v_{1}-v_{2}$, whereas demanding the object only yields
$o_{1} = v_{1} - p$.  Note that here it is essential that $v_{2} > 0$,
since it allows us to conclude that also $p-1 > 0$, i.e., we have not
reached the last auction round yet, there will be another round with a
lower price.

Denote the new precondition as $Inv_{3}$.  Note that $Inv \wedge w=0
\subseteq Inv_{3}$.  Hence, by using the composition rule and the
logical consequence rule, we have established that $Inv$ is indeed an
invariant of the loop.  Hence, we can apply the while rule to derive
that ${\mathcal I} \vdash$
\begin{tabbing}
$\{Inv\}$ \\
$\mathtt{while}$ \= $\: p > 0 \wedge w=0 \: \mathtt{do}$ \\ \>
$\mathtt{ch}_{\{1,2\}}(\{x_1,x_2\})$; \\ \>
    $\mathtt{if} \: x_1 > 0 \:$ \= $\mathtt{then} \: w:=1$ \\ \>
    \> $\mathtt{else} \: \mathtt{if} \: x_2 > 0 \:$ \= $\mathtt{then}
    \: w:=2$ \\ \>
    \>\> $\mathtt{else} \: p:=p-1$ \\
$\{Inv \wedge \neg (p>0 \wedge w=0)\}$
\end{tabbing}

So to conclude the verification of the Dutch auction, it suffices to
note two things.  First, $Inv \wedge \neg (p>0 \wedge w=0)
\subseteq Q$, and hence we can apply the logical consequence rule to
obtain the desired postcondition $Q$.  Second, we have ${\mathcal I}
\vdash$
\begin{tabbing}
$\{v_{1} \geq v_{2} > 0 \wedge init \geq v_{2} \wedge o_{1} = v_{1}
- v_{2} \wedge o_{2} = 0 \}$ \\
$p:=init$; \\ $w:=0$ \\
\{Inv\}
\end{tabbing}
Hence, using the composition rule, we have now shown that ${\mathcal
I} \vdash$
\begin{tabbing}
$\{v_{1} \geq v_{2} > 0 \wedge init \geq v_{2} \wedge o_{1} = v_{1} -
v_{2} \wedge o_{2} = 0 \}$ \\
$p:=init$; \\ $w:=0$; \\
$\mathtt{while}$ \= $\: p > 0 \wedge w=0 \: \mathtt{do}$ \\ \>
$\mathtt{ch}_{\{1,2\}}(\{x_1,x_2\})$; \\
    \> $\mathtt{if} \: x_1 > 0 \:$ \= $\mathtt{then} \: w:=1$ \\
    \>\> $\mathtt{else} \: \mathtt{if} \: x_2 > 0 \:$ \= $\mathtt{then}
    \: w:=2$ \\
    \>\>\> $\mathtt{else} \: p:=p-1$ \\
$\{(w=1 \rightarrow (o_1=v_1-p \wedge o_2=0)) \wedge (w=2
\rightarrow (o_1=0 \wedge o_2=v_2-p))$ \\ $\wedge \: (w=0 \rightarrow
(o_1=0 \wedge o_2=0))\}$
\end{tabbing}

Note that the verification process has revealed two crucial details
which had to be added to our original precondition $P$.  First, $init
\geq v_2$.  This means that we need to make sure that we start the
auction at a price that is high enough.  If the players' valuations
are not known, the choice of the initial price can indeed be a
problem.  On the other hand, the condition tells us exactly what
``high enough'' means, in particular, the initial price does not need
to exceed everybody's valuation.  Second, $v_2 > 0$.  Hence, it does not
suffice if only a single player has a non-zero valuation of the
object.  The problem here lies in the fact that in order to obtain the
object one has to pay at least something, and if the other player's
valuation is zero, that something is more than the other player's
valuation, and hence the payoff is in turn lower than expected.  Hence,
we have succeeded in verifying that $(\alpha,Q)$ does indeed implement
the social choice correspondence of section \ref{auctions} associated
with the second-price auction, on condition that $init \geq v_{2} >
0$.

Finally, it should be emphasised again that the weak equivalence of the Dutch
auction and the sealed-bid second-price auction demonstrated here is very
weak indeed, since these auctions are very different.
Crucially, in the sealed-bid second-price
auction, a player does not need to know the other player's valuation. 
It suffices that each player submits his own valuation as a bid.  In
the Dutch auction, however, obtaining the same equilibrium outcome
requires the player with the higher valuation to know the valuation of
the other player so that he can decide to shout out just at the right
moment.  Hence, the two auctions do not satisfy the same knowledge
preconditions. The standard result concerning the equivalence between Dutch 
auction and first-price auction does take these knowledge preconditions into 
account.

%
%
\section{Conclusions}
%
%

Two main directions for future research present themselves: On the
foundational side, the question arises whether the present
approach can also be applied to other equilibrium notions. We have
already remarked that while the calculus presented can also be
used to reason about Nash equilibria, the non-compositional nature
of these equilibria stands in the way of a complete calculus.
Hence, alternative equilibrium notions that promise to be amenable
to our approach will be refinements of subgame-perfect equilibria.
Second, we mentioned already that an intensional approach to pre-
and postconditions is worth developing. For this, the crucial
question is whether the logic used (FOL) and the expressiveness
results obtained for programs can be carried over to mechanisms.

At the most general level, we hope that this paper has shown that
tools from computational logic can be extended from program
verification to the verification of game-theoretic mechanisms. The
examples provided should suffice to convince the reader of the
variety of possible applications of such an extension. The
semantics of the correctness assertions for mechanisms is more
complex than for programs, but this is counterbalanced by the fact
that the mechanisms we would like to verify (e.g., spectrum
auctions for telecommunication markets) may turn out to be simpler
than their counterparts in computer software (e.g., operating
systems).

%
\subsection*{Acknowledgments}
%
For their comments and suggestions, I would like to thank the anonymous referees,
Peter McBurney, Mike Wooldridge and the members of Rohit Parikh's seminar 
at the Graduate Center of the City University of New York (CUNY).

\bibliographystyle{plain}
\bibliography{marc}

\begin{thebibliography}{10}

\bibitem{Apt-Olderog}
K.R. Apt and E.R. Olderog.
\newblock {\em Verification of Sequential and Concurrent Programs}.
\newblock Springer, second edition, 1997.

\bibitem{Brams:voting}
S.~Brams and P.C. Fishburn.
\newblock Voting procedures.
\newblock In K.~Arrow, A.~Sen, and K.~Suzumura, editors, {\em Handbook of
  Social Choice and Welfare}, volume~1. North-Holland, 2002.

\bibitem{alternating-TM}
A.~Chandra, D.~Kozen, and L.~Stockmeyer.
\newblock Alternation.
\newblock {\em Journal of the ACM}, 28(1):114--133, 1981.

\bibitem{Clarke}
E.~Clarke, O.~Grumberg, and D.~Peled.
\newblock {\em Model Checking}.
\newblock MIT Press, 1999.

\bibitem{D}
E.~Dijkstra.
\newblock {\em A Discipline of Programming}.
\newblock Prentice-Hall, 1976.

\bibitem{Francez}
N.~Francez.
\newblock {\em Program Verification}.
\newblock Addison-Wesley, 1992.

\bibitem{Hoare-logic}
C.A.R. Hoare.
\newblock An axiomatic basis for computer programming.
\newblock {\em Communications of the ACM}, 12(10):576--580, 1969.

\bibitem{Kraus}
S.~Kraus.
\newblock {\em Strategic Negotiation in Multiagent Environments}.
\newblock MIT Press, 2001.

\bibitem{Moore:implementation}
J.~Moore.
\newblock Implementation, contracts, and renegotiation in environments with
  complete information.
\newblock In J.-J. Laffont, editor, {\em Advances in Economic Theory: Sixth
  World Congress}, volume~1. Cambridge University Press, 1992.

\bibitem{Nielson}
H.~R. Nielson and F.~Nielson.
\newblock {\em Semantics with Applications}.
\newblock Wiley, 1992.

\bibitem{Osborne}
M.~Osborne and A.~Rubinstein.
\newblock {\em A Course in Game Theory}.
\newblock MIT Press, 1994.

\bibitem{Parikh}
R.~Parikh.
\newblock The logic of games and its applications.
\newblock In M.~Karpinski and J.~van Leeuwen, editors, {\em Topics in the
  Theory of Computation}, Annals of Discrete Mathematics 24. Elsevier, 1985.

\bibitem{Parikh:SoSo}
R.~Parikh.
\newblock Social software.
\newblock {\em Synthese}, 132(3):187--211, 2002.

\bibitem{coal1}
M.~Pauly.
\newblock A modal logic for coalitional power in games.
\newblock {\em Journal of Logic and Computation}, 12(1):149--166, 2002.

\bibitem{RoE}
J.S. Rosenschein and G.~Zlotkin.
\newblock {\em Rules of Encounter: Designing Conventions for Automated
  Negotiation Among Computers}.
\newblock MIT Press, 1994.

\end{thebibliography}

\end{document}